\documentclass{article}
\title{String effects in Polyakov loop correlators}
\author{M. Caselle$^a$, M.Panero$^a$ and P. Provero$^{b,a}$}
\date{April 2002}
\usepackage{amssymb}
\usepackage{graphicx}
\usepackage{latexsym}
\newcommand{\eq}{\begin{equation}}
\newcommand{\qe}{\end{equation}}
\newcommand{\ear}{\begin{eqnarray}}
\newcommand{\rae}{\end{eqnarray}}
\newcommand{\Z}{\mathbb{Z}}
\begin{document}
\maketitle
\noindent
$^a$ Dipartimento di Fisica Teorica, Universit\`a di Torino and INFN,
sezione di Torino, via P. Giuria, 1, I-10125 Torino, Italy.
\vskip0.5cm\noindent
$^b$ Dipartimento di Scienze e Tecnologie Avanzate, Universit\`a del
Piemonte Orientale ``A. Avogadro'', and INFN, gruppo collegato di
Alessandria, Corso Borsalino 54, I-15100 Alessandria, Italy.
\vskip0.5cm\noindent
e-mail: caselle, panero, provero@to.infn.it
\begin{abstract}\noindent
We compare the predictions of the effective string description of
confinement in finite temperature gauge theories to high precision
Monte Carlo data for the three-dimensional $\Z_2$ gauge theory. First
we review the predictions of the free bosonic string model and their
asymptotic behavior in the various regimes of physical interest. Then
we show that very good agreement with the Monte Carlo data is
obtained, for temperatures not too close to the deconfinement one
(typically $T\le T_c/3$). For higher temperatures, higher order
effects are not negligible: we show that they are accurately modeled
by assuming a Nambu-Goto string action and computing its partition
function at next-to-leading order.
\end{abstract}
\section{Introduction}
Two color sources in a confining
gauge theory are bound together by a thin flux tube, which 
can fluctuate like a massless string: this hypothesis
constitutes the core of the effective string description of
confinement. 
Besides providing an appealing, qualitative explanation for the linear
behavior of the confining potential at large distances, this picture
provides us with quantitative, testable predictions about this highly
non perturbative regime of gauge theories, as pointed out  in the
seminal papers 
by L\"uscher, Symanzik and Weisz \cite{Luscher:1980fr, Luscher:1980ac}:
the {\em quantum
fluctuations} of the flux tube produce measurable effects on the gauge
invariant correlation functions of the gauge theory, namely Wilson
loops and  Polyakov loop correlation functions. 
\par
Among these effects, the most widely known is the L\"uscher
term correction to the zero-temperature linear confining potential:
\eq
V(R)\sim\sigma R - \frac{c\pi}{24 R}
\label{luscher}
\qe
where $c$ is the conformal anomaly of the two-dimensional field theory
describing 
the flux tube fluctuations: $c=d-2$ for a free bosonic string in a
$d$-dimensional spacetime.  
\par
The predictive power of the effective string picture goes well beyond
Eq.(\ref{luscher}), as it extends to the full functional form of the
Wilson loop expectation values at large distances:
\eq
<W(R,T)>=e^{-\sigma RT+p(R+T)+k}Z_{q}(R,T)
\label{full_wilson}
\qe
where  $Z_{q}(R,T)$ is the partition function of the two-dimensional
quantum field theory describing the quantum fluctuations of the flux
tube. For the free bosonic string, this is simply the theory of $d-2$
free massless scalar fields living on the rectangle defined by the Wilson
loop:
\eq
Z_{q}(R,T)\propto\left[\frac{\eta(\tau)}{\sqrt{R}}
\right]^{-\frac{d-2}{2}}
\label{zetaq}
\qe  
where $\eta(\tau)$ is the Dedekind $\eta$ function and
$\tau=iT/R$. The prediction Eq.(\ref{zetaq}) was successfully compared
to Monte Carlo data for three-dimensional $\Z_2$ gauge theory in
Ref.\cite{Caselle:1996ii}, thus confirming the validity of the string
description, and showing its full predictive power.
\par
It must be kept in mind that the free string description that
gives Eq.(\ref{zetaq}) is an effective description that is expected to
hold in the {\em long distance} regime. 
The ``true'' string theory
describing gauge theories at all length scales, if it exists at all,
certainly contains complicated string self-interactions, probably
described by non-renormalizable terms in  the string action.  
Precisely because they are non-renormalizable, however, these terms
should become negligible in 
the infrared limit, so that the free string model is expected to
describe the physics of confinement for large
distances independently of the details of the full, interacting theory.
Among the results of
Ref.\cite{Caselle:1996ii} is a quantitative estimate of the typical
distances, in physical units, where the free string picture becomes
numerically accurate: if the physical
distances studied are too small, one cannot unambiguously determine
which string model actually describes the flux tube fluctuations\footnote{This
is the reason why earlier studies, probing shorter physical distances
due to the smaller computational power available, could not identify
the free bosonic string as the correct model, and actually suggested a
fermionic string model as the best description of the confining string
both at zero \cite{Caselle:1990qg} and finite temperature
\cite{Caselle:1992ga}.}. 
\par
In this paper we consider the effective string description for a
confining gauge theory at finite temperature. The gauge invariant quantity
of interest is now the correlation function of two Polyakov loops.
The effective string description predicts its dependence on the
temperature and on the distance between the two loops. It is this
prediction that we want to compare to Monte Carlo data. We chose to
work again with the three-dimensional $\Z_2$ gauge theory, where
very high precision can be achieved on large lattices, to make
our test as stringent as possible. 
\par
Also in the finite-temperature case it is important to keep in mind
the fact that the free string picture is an effective description
valid in the infrared limit, and to determine quantitatively the
distance and temperature scales where it becomes numerically accurate.
First of all, one has to take into account the known fact that the
actual color flux tube has a finite thickness of the order $\sim
1/T_c$, where $T_c$ is the deconfinement critical temperature
\cite{Caselle:1995fh}: 
therefore the free string description, which is based on an
idealized, one-dimensional flux tube,  certainly must break
down at distances lower that $1/T_c$. Moreover, we will show that
the string self-interaction effects mentioned above cannot be
neglected for temperatures close to $T_c$: within the
accuracy of our data, one has to take $T\le T_c/3$ to find good
agreement between Monte Carlo data and free string predictions. 
\par
We will also show that by modeling the self-interaction of the
confining string by a Nambu-Goto action and computing the string
partition function at next-to-leading order one can successfully take
into account the most important corrections to the free string
picture: one then obtains good agreement with the Monte Carlo data for
temperatures up to $\sim T_c/2$ (without introducing any new
adjustable parameters).
\par
Since the effective string description is believed to be universal,
that is to hold for all confining gauge theories irrespective of the
gauge group or the space-time dimensionality, 
our conclusions are likely to apply also
to $SU(N)$ gauge theories in $2+1$ and
$3+1$ dimensions. Agreement between free string theory predictions and
lattice determinations of the interquark potential in $SU(N)$ gauge
theories was recently reported by various groups
\cite{Necco:2001xg, Lucini:2001nv, Gliozzi:1999wq}. 
\section{Effective string predictions}
The quantum fluctuations of the flux tube at finite temperature are
described, in the free string model, by $d-2$ free massless scalar
fields obeying Dirichlet boundary conditions on the two Polyakov loops
and periodic boundary conditions in the compactified  direction
(euclidean time).
The correlation of two Polyakov loops at distance $R$ is then
predicted to be
\eq
\langle P(0)P^{\dagger}(R)\rangle=\exp\left[-F(R,L)\right]
\qe
where the free energy $F(R,L)$ depends on the inverse temperature
$L\equiv 1/T$ ({\it i.e.} the lattice size in the time direction) and
the distance $R$, and is given by a classical and a quantum
contribution:
\eq
F(R,L)=F_{\rm cl}(R,L)+F_{\rm q}(R,L)
\qe
\par
The classical term corresponds to the area law: 
\eq
F_{\rm cl}(R,L)=\sigma_0 L R +k(L)
\label{area}
\qe
Since the difference between Wilson loops and Polyakov loop
correlation functions is, in this picture, completely taken into
account by the different choices of boundary conditions for the string
fluctuations, $\sigma_0$ is the
zero-temperature string tension, extracted, say, from the Wilson loop
at the same value of the gauge coupling.
\par
The quantum term encodes the flux tube fluctuations, and is
equal to the free energy of the $d-2$ massless scalar fields
describing them\cite{Minami:1977hd,deForcrand:1984cz,Flensburg:wu}:
\eq
F_{\rm q}(R,L)=(d-2)\log\eta(\tau)\ \ \ \ \tau\equiv\frac{iL}{2R}  
\label{fq}
\qe
where $\eta$ is again the Dedekind eta function:
\eq
\eta(\tau)=q^{1\over24}\prod_{n=1}^\infty(1-q^n)\hskip0.5cm
;\hskip0.5cmq=e^{2\pi i\tau}~~~,\label{eta}
\qe
According to the value of the ratio $-i\tau=L/2R$ one can use the two
expansions:
\begin{description}
\item{$2R<L$}
\eq
F_q(R,L)=(d-2)\left[-\frac{\pi L}{24 R}
+\sum_{n=1}^\infty \log (1-e^{-\pi nL/R})\right]
\label{zsmalltot}
\qe
\item{$2R>L$}
\eq
F_q(R,L)=(d-2)\left[-\frac{\pi R}{6 L}+\frac{1}{2} \log\frac{2R}{L}
+\sum_{n=1}^\infty \log (1-e^{-4\pi nR/L})\right]
\label{zbigtot}
\qe
\end{description}
The latter expression shows that the Polyakov loop correlation
function should decay at large $R$ as
\eq
\langle P(0)P^{\dagger}(R)\rangle\sim\exp \left[-\frac{\sigma(T)}{T}
R\right] 
\qe
with a temperature-dependent string tension $\sigma(T)$ given by
\eq
\sigma(T)=\sigma_0-(d-2)\frac{\pi T^2}{6}
\label{sigmat}
\qe
\section{Comparison with Monte Carlo data}
\subsection{At what scales is the free string picture expected to hold?}
To compare the free string predictions described in the previous
section to Monte Carlo data, we need to establish in which regime of
distance $R$ and temperature $1/L$ we expect these predictions to be
fulfilled. 
\par
It is known that the picture of a
one-dimensional confining string is an idealization valid for long
distances. The flux tube has actually a finite thickness of order
$1/T_c$, where $T_c$ is the deconfinement critical temperature
\cite{Caselle:1995fh}. 
Therefore we expect the free string picture to be accurate for $R>1/T_c$. 
\par
We also know that the free string picture must break down for
temperatures close to $T_c$: in fact if we were to believe the free
string picture for all temperatures up to $T_c$,
Eq.(\ref{sigmat}) would predict the value of the latter to be
$T_c=\sqrt{6\sigma_0/\pi}$, a prediction that turns out to be very far
from the true value.
\par
A much better prediction for the
dimensionless ratio $T_c^2/\sigma_0$ is obtained if one assumes for
the confining 
string a Nambu-Goto action, proportional to the area of the surface
spanned by the string. One can then predict
\cite{Pisarski:1982cn,Olesen:1985ej} 
\eq
\sigma(T)=\sigma_0\sqrt{1-\left(\frac{T}{T_c}\right)^2}
\label{olesen}
\qe
and
\eq
T_c^2=\frac{3\sigma_0}{(d-2)\pi}
\label{olesen2}
\qe
which reduces to Eq.(\ref{sigmat}) for $T<<T_c$. 
The
impressive agreement one finds with Monte Carlo data for various gauge
groups suggests that these equations give at least a good
approximation to the true behavior.
\par
Therefore we can use Eq.(\ref{olesen}) to estimate the range of
temperatures in which string interaction effects can be safely
neglected and the free string picture is expected to be accurate:
expanding the square root in Eq.(\ref{olesen}) to next-to-leading order
we find
\eq
\sigma(T)\sim \sigma_0
\left[1-\frac12\left(\frac{T}{T_c}\right)^2
-\frac18\left(\frac{T}{T_c}\right)^4\right] 
\qe
We expect the free string picture to give an accurate description of
the data when the last term in parentheses is comparable to the
accuracy in the determination of $\sigma_0$. In our case such accuracy
is of order 1\%, so that we do not expect to be able to use the free string
prediction Eq.(\ref{sigmat}) for temperatures higher than $\sim T_c/2$. We will
see below that this estimate turns out to be too optimistic, since
good agreement between free string and Monte Carlo data is obtained
only up to $\sim T_c/3$. Note however that this determination of the
limiting temperature ratio $T/T_c$ for the free string picture
depends in an essential way from the accuracy of the Monte Carlo data,
and hence has no intrinsic meaning.
\subsection{Monte Carlo simulation}
We simulated the three-dimensional
$\Z_2$ gauge model with Wilson action:
\eq
S=-\sum_\Box \sigma_1 \sigma_2 \sigma_3 \sigma_4
\qe
where the sum is extended to all plaquettes of a cubic lattice, and
$\sigma_1,\dots,\sigma_4$ are $\Z_2$ variables defined on the four
links around the plaquette. The partition function is
\eq
Z(\beta)=\sum_{\{\sigma\}}\exp -\beta S
\qe
where the coupling $\beta$ is the same for all directions. 
\par
With these conventions the model, at zero temperature,  
is known to undergo a {\em
roughening} transition at $\beta_r=0.4745(2)$, where the strong
coupling expansion of the Wilson loop ceases to 
converge, signaling the fact that the flux tube fluctuations become
massless: for $\beta>\beta_r$ we expect the free bosonic string
description to hold. Increasing $\beta$ we find a (bulk) {\em
deconfinement} transition at $\beta_c=0.76141337(22)$\cite{Blote:1999cf}
where the string 
tension extracted from Wilson loops vanishes. 
\par
At finite temperature, the model undergoes a deconfinement transition,
at a coupling $\beta_d(L)$ that depend
on the inverse temperature $L\equiv 1/T$. For several values of $L$ 
the critical coupling is known to high precision.
In particular we have \cite{Caselle:1995wn} 
\ear
\beta_d(4)&=&0.73107(2)\label{bd4}\\
\beta_d(6)&=&0.746035(8)\label{bd6}
\rae
\par
In our simulations we used the gauge version of the microcanonical
demon algorithm \cite{Creutz:1983ra,Bhanot:1983bi} combined with the
canonical update of the demons \cite{Rummukainen:1992fc}, described in
detail in Ref.~\cite{Agostini:1996xy}, and implemented, as in
Ref.~\cite{Caselle:1996ii}, in multi-spin coding technique.
We fixed the coupling $\beta$ to coincide with one of the values reported
in Eqs. (\ref{bd4},\ref{bd6}), so as to be able to control 
the value of $T/T_c$ by varying
the lattice size $L$. The lattice size in the two space-like directions
was fixed at $N_s=48$, corresponding to at least ten times the
correlation length in the space-like directions. 
\par
For these two values of $\beta$  we simulated the
model at several values of the inverse temperature $L$, all
corresponding to temperatures up to $T_c/2$. In Tab. 2 we show
for each of the two $\beta$ values the values of $L$ we used in the
simulations, and the value of the zero-temperature string tension
$\sigma_0(\beta)$. The latter were evaluated by interpolating the
high-precision estimates published in
Refs.\cite{Hasenbusch:1992zz,Caselle:1994df,Hasenbusch:1997} 
for several reference
values of $\beta$ with the scaling formula
\eq
\sigma_0(\beta)=
\sigma_0(\beta_{ref}) 
\left(\frac{\beta-\beta_c}{\beta_{ref}-\beta_c}\right)^{2\nu}
\qe
where $\beta_{ref}$ is the closest value for which a direct Monte
Carlo estimate is available, and $\nu$ is the correlation length
critical index. Neglecting corrections to scaling introduces a
systematic error that can be estimated by evaluating $\sigma_0(\beta)$
using two nearby reference values. The errors quoted in Tab. 2 are the
sum of the statistical and systematic errors.
\begin{table}[h]
\begin{center}
\begin{tabular}{|c|c|c|c|}
\hline
$\beta$&$1/T_c$&$L$&$\sigma_0 a^2$\\
\hline
0.73107&4&8-14&0.0440(3)\\
\hline
0.74603&6&12,14,15,16,17,18,20&0.018943(32)\\
\hline
\end{tabular}
\end{center}
\caption{\sl Temporal lattice sizes and zero-temperature string
tension for our values of $\beta$.} 
\end{table}
\par
For each $\beta$ and $L$ in Tab. 2 we 
evaluated the ratio
\eq
Q(R)\equiv \ln \frac{G(R)}{G(R+1)}
\qe
where $G(R)$ is the Polyakov loop correlation function at distance
$R$:
\eq
G(R)=\langle P(x) P^{\dagger}(x+R)\rangle
\qe
for all $R$ between 0 and $N_s/2=24$. Errors on $Q(R)$ were evaluated
using a standard jackknife procedure. 
\par
From Eq.(\ref{area}) we see that if the quantum fluctuations of the
confining string are neglected, $Q(R)$ is predicted to be a constant:
\eq
Q(R)=\sigma_0(\beta) L
\qe
so that the difference
\eq
Q_q(R)\equiv Q(R)-\sigma_0(\beta) L
\label{Qq}
\qe
can be interpreted as the quantum contribution: the free string
prediction for this quantity is 
\eq
Q_q(R)=F_q(R+1,L)-F_q(R,L)
\label{Qqfree}
\qe
with $F_q$ given by Eq.(\ref{fq}). 
\par
The complete Monte Caro results will be presented in a forthcoming
publication, together with details about the data analysis
procedure. Here we limit ourselves to a smaller sample of data,
sufficient to evidentiate the most important results of the analysis,
that are likely to be relevant also for $SU(N)$ gauge theories.
\subsection{Results at $T\le T_c/3$: the free string.}
All the data at  $T\le T_c/3$ are in good agreement with the free
string predictions: in Figs. 1-2 we have plotted the MC values of
$Q_q(R)$ defined by Eq.(\ref{Qq}) and the free string prediction Eq.
(\ref{Qqfree}) for our two $\beta$ values at $T=T_c/3$. 
The data are plotted as a function of 
\eq
z=\frac{i}{\tau}=\frac{2R}{L}\ \ ,
\qe
which is the ``natural'' variable for the effective string prediction.
In the three figures we have
shown only the data with $R>1/T_c$, since for lower $R$ we do not
expect the free string to describe the data. Moreover we have not
shown the data at very large $R$, where the statistical error on $Q_q$
becomes larger than the effect we are studying.
\begin{figure}
\centering
{\includegraphics[width=12.cm]{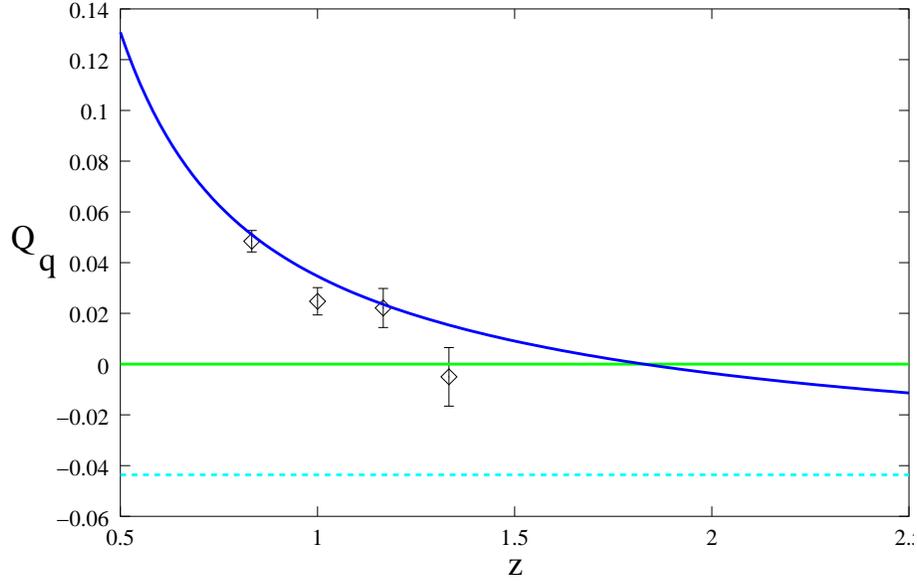}}
\caption{\sl Comparison between free string prediction and Monte Carlo
data for $Q_q(R)$ as defined in Eq.(\ref{Qq}). The Monte Carlo data
are taken at $\beta=0.73107$ and $T=T_c/3=(12 a)^{-1}$, and are
plotted as a function of $z=2R/L$. The solid line is the
prediction Eq.(\ref{Qqfree}) for this quantity. The dashed line is the
asymptotic value $-\pi/6L$ of the free string prediction for large $z$.}
\end{figure}
\begin{figure}
\centering
{\includegraphics[width=12.cm]{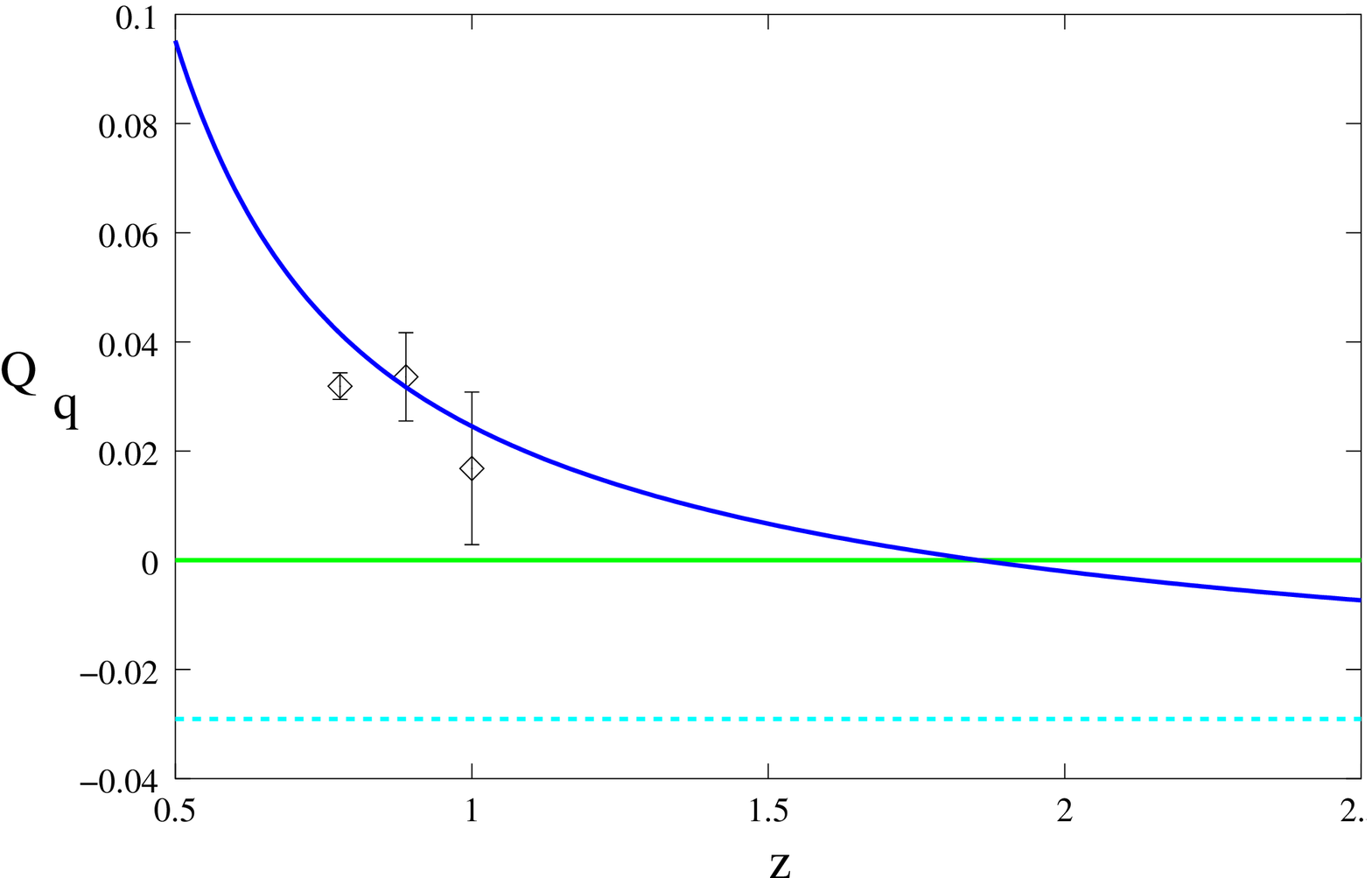}}
\caption{\sl Same as Fig. 1 for $\beta=0.74603$ and $T=T_c/3=(18 a)^{-1}$}
\end{figure}
\par
A few interesting things can be learned by looking at these figures:
\begin{itemize}
\item
The data we can use for the comparison are all in the region $z\sim 1$. This is
unavoidable since small values of $z$ correspond to distances smaller
than the width of the flux tube, as explained above, while data at
large $z$ have large statistical uncertainties.
\item
In this region the free string prediction is very far from its
asymptotic behavior for large $R$, where it approaches the constant
\eq
F_q(R+1,L)-F_q(R,L)\to -\frac{(d-2)\pi}{6L}
\qe
shown for comparison in the figures: the string correction for $z\sim
1$ has actually the opposite sign with respect to its asymptotic
behavior.
\item
The function $F_q(z\equiv 2R/L)$ has a zero in $z\sim 1.8$. This is
due to the fact that the asymptotic term $\pi/(6L)$ and the
subdominant logarithmic term that appears in Eq.(\ref{zbigtot}) have
opposite signs. This in turns makes the string corrections much harder
to detect in the Polyakov loop correlation functions than in the
Wilson loop case, where no such cancellation occurs.
\end{itemize}  
\subsection{Results at $T\sim T_c/2$: the interacting string.}
If the same comparison between Monte Carlo data and free string
predictions is performed at temperatures closer to the deconfinement
one, significant discrepancies begin to appear, showing that
the free string contribution is not
sufficient to account for the quantum fluctuations of the flux tube at
such temperatures: string interaction effects, or more precisely
the self-interaction of the world-sheet fields describing the string
configuration, become non
negligible. 
\par
The simplest way to estimate the effect of these self-interactions is
to assume a Nambu-Goto action for the effective string. On one hand,
this is exactly the assumption that leads one to the remarkably
successful predictions Eqs.(\ref{olesen},\ref{olesen2}). On the
other hand this same choice of string action accurately describes the
string interaction effects for fluctuating interfaces in
three-dimensional statistical models\cite{Caselle:1994df,Provero:fd}.
\par
From this assumption, one can compute the string partition function to
next-to-leading order in the dimensionless expansion parameter
$(\sigma R L)^{-1}$, with the choice of boundary conditions relevant to
the Polyakov loop correlation function, namely Dirichlet on the two loops
and periodic in the Euclidean time direction. This calculation was
performed in Ref. \cite{Dietz:1983uc} using $\zeta$-function regularization; it can be
shown \cite{Caselle:1996kd} that the same result is obtained with several other choices of
regularization. 
\par
Including these contribution the string free energy Eq.(\ref{fq})
becomes
\eq
F_q^{(NLO)}(R,L)=(d-2)\left[\log\eta(\tau)-\frac{\pi^2 L}{1152\ \sigma
R^3}\left[2
E_4(\tau)-E_2^2(\tau)\right]\right]+O\left(\frac{1}{\left(\sigma L
R\right)^2}\right) 
\label{nlo}
\qe
where $E_2$ and $E_4$ are the Eisenstein functions. The latter can be
expressed in power series:
\ear
E_2(\tau)&=&1-24\sum_{n=1}^\infty \sigma(n) q^n\\
E_4(\tau)&=&1+240\sum_{n=1}^\infty \sigma_3(n) q^n\\
q&\equiv& e^{2\pi i\tau}
\rae 
where $\sigma(n)$ and $\sigma_3(n)$ are, respectively, the sum of all
divisors of $n$ (including 1 and $n$), and the sum of their cubes.
The modular transformation properties
\ear
E_2(\tau)&=&-\left(\frac{i}{\tau}\right)^2
E_2\left(-\frac1\tau\right) +\frac{6i}{\pi\tau}\\
E_4(\tau)&=&\left(\frac{i}{\tau}\right)^4
E_4\left(-\frac1\tau\right)
\rae
are also useful.
Note that the inclusion of next-to-leading terms does not require the
introduction of any new free parameter, so that the predictive power
is the same as for the free string case.
\par
In Figs. 3-4 we show the Monte Carlo data for the string fluctuation
contribution defined in Eq.(\ref{Qq}) for $T=T_c/2$ compared to both the free
string prediction and the Nambu-Goto string at
next-to-leading order Eq.(\ref{nlo}). We can see that Eq.(\ref{nlo})
accurately describes the deviations of the data from the free string
picture. 
\begin{figure}
\centering
{\includegraphics[width=12.cm]{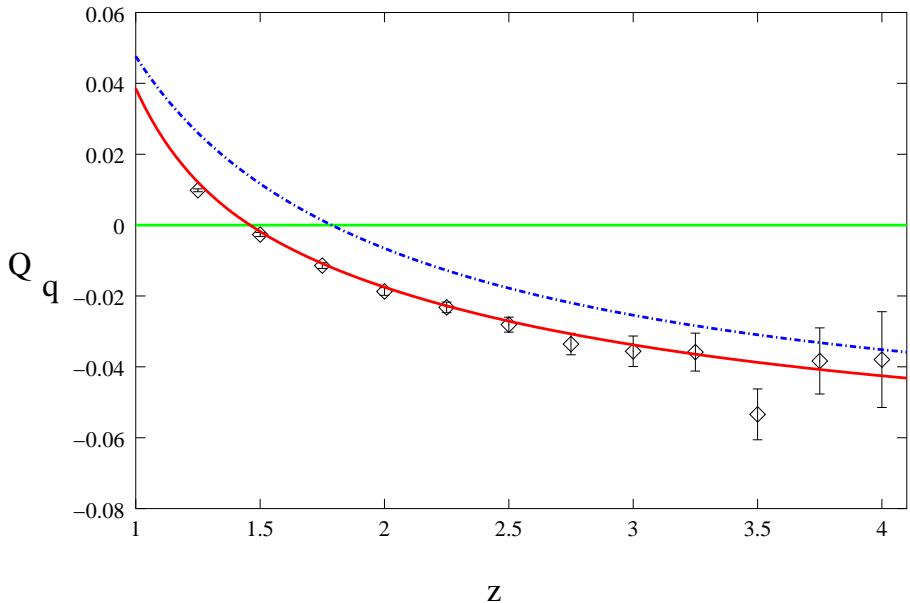}}
\caption{\sl Data for $Q_q$ at  $\beta=0.73107$ and $T=T_c/2=(8
a)^{-1}$. The 
solid line is the full NLO string prediction, while the dash-dotted
line is the free string prediction.} 
\end{figure}
\begin{figure}
\centering
{\includegraphics[width=12.cm]{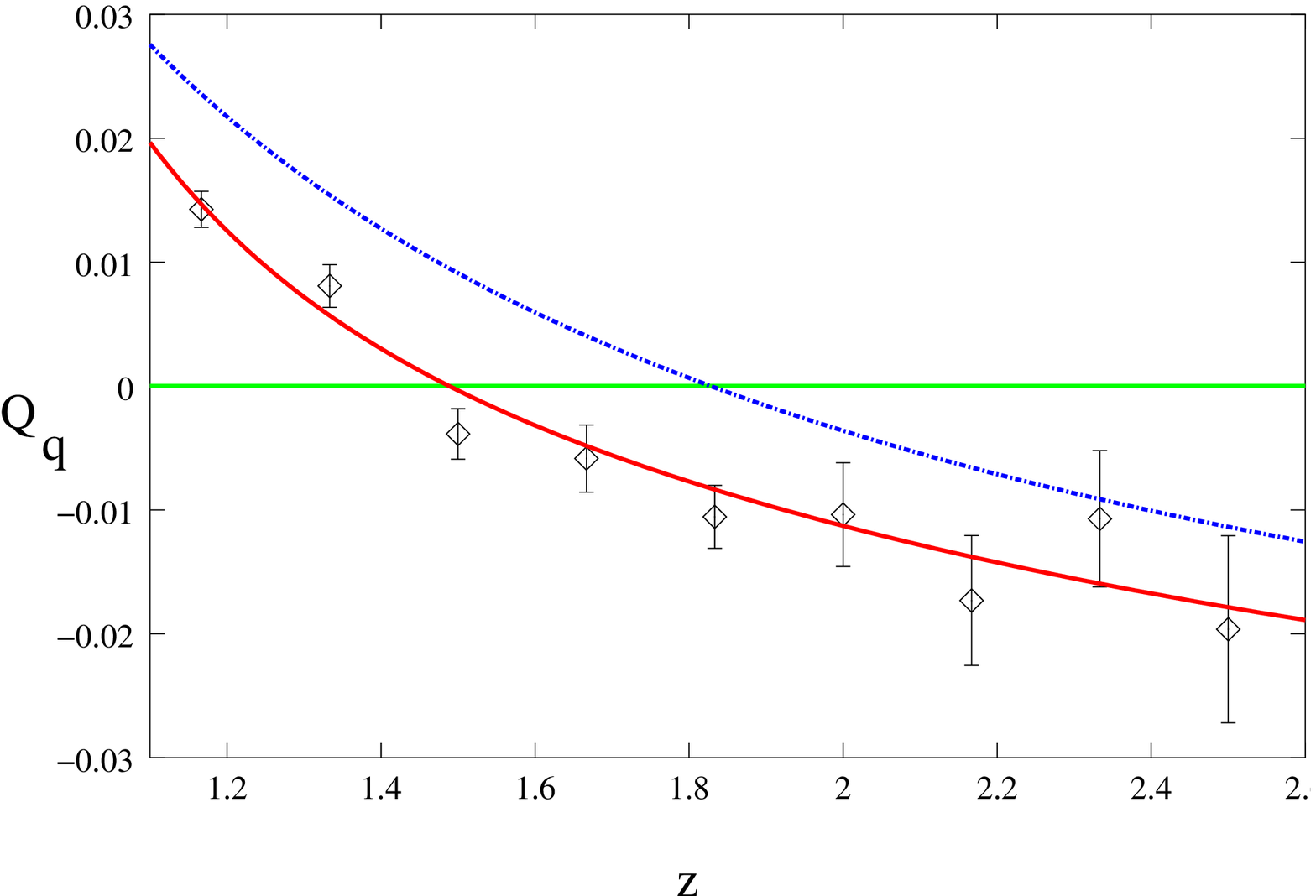}}
\caption{\sl Same as Fig. 4 for $\beta=0.74603$ and $T=T_c/2=(12 a)^{-1}$} 
\end{figure}
\par
A word of caution is however in order: while the free string
prediction Eq.(\ref{fq}) is the universal infrared
limit of a large class of interacting string models (see {\em e.g.}
the various  models considered in \cite{Dietz:1983uc}), the
interaction contributions depend on the choice of the model. Therefore
while the validity of the free string model in the infrared limit
$R,L\to\infty$ is basically a consequence of the
masslessness and bosonic character of the fields describing the string
fluctuations, the choice of the Nambu-Goto action  
is a much stronger statement, that cannot as far as we
know be based on simple physical grounds. Nevertheless, it is certainly
noteworthy that the same choice of action gives accurate predictions
for fluctuating interfaces \cite{Caselle:1994df,Provero:fd} and
Polyakov loop correlation functions.
\section{Conclusions}
The study of Polyakov loop correlation functions in the
three-dimensional $\Z_2$ model allows us to gain some insight on the
effect of the fluctuations of the confining flux tube, which is likely
to be relevant also in more complicated and realistic gauge theories. The
main points we have elucidated are:
\begin{enumerate}
\item
The quantum fluctuations of the confining string give a computable and
numerically relevant contribution to the Polyakov loop correlation
functions in the confined phase.
\item
The free bosonic string model describes well such contributions provided the
temperature $T$ is not too close to the critical one: in our case
$T\le T_c/3$ is required to find good agreement.
\item
When the temperature gets closer to $T_c$, next-to-leading effects in
the string partition function become important. Assuming a Nambu-Goto
action for the confining string one can compute these effects and find
excellent agreement with the Monte Carlo data also at $T\sim T_c/2$.
\end{enumerate}
\vskip1.cm
\noindent
{\bf Acknowledgement}
\vskip0.5cm\noindent
We would like to thank M. Hasenbusch for reading a previous version of this 
work and offering several useful suggestions.

\end{document}